\documentclass[useAMS,usenatbib]{mn2e}
\usepackage{graphicx}
\def\new#1{{#1}}

\title[Fast infrared variability from a relativistic jet]{Fast infrared variability from a relativistic jet in GX 339--4\thanks{Based on observations made with ESO Telescopes at the Paranal 
Observatory under programme ID 281.D-5034}}
\author[P. Casella et al.]{P. Casella$^{1,2}$\thanks{E-mail:
p.casella@soton.ac.uk},
T. J. Maccarone$^{2}$, K. O'Brien$^{3}$, R. P. Fender$^{2}$, D. M. Russell$^{1}$, 
\newauthor M. van der Klis$^{1}$, A. Pe'er$^{4}$, D. Maitra$^{1}$, D. Altamirano$^{1}$, T. Belloni$^{5}$, G. Kanbach$^{6}$,
\newauthor M. Klein-Wolt$^{7}$, E. Mason$^{3}$, P. Soleri$^{1}$, A. Stefanescu$^{6,8,9}$
, K. Wiersema$^{10}$, R. Wijnands$^{1}$\\
$^{1}$Astronomical Institute "A. Pannekoek", University of Amsterdam, Science Park 904, 1098 XH, Amsterdam, The Netherlands\\
$^{2}$School of Physics and Astronomy, University of Southampton, Southampton, Hampshire, SO17 1BJ,
UK\\
$^{3}$European Southern Observatory, Casilla 19001, Santiago 19, Chile\\
$^{4}$Space Telescope Science Institute, 3700 San Martin Dr., Baltimore, MD 21218, USA\\
$^{5}$INAF-Osservatorio Astronomico di Brera, Via E. Bianchi 46, I-23807 Merate (LC), Italy\\
$^{6}$Max-Planck Institut f\"ur Extraterrestrische Physik, 85741 Garching bei M\"unchen, Germany\\
$^{7}$Altran BV., Hendrik Walaardt Sacrestraat, 405, 1117 BM, Schiphol Oost, The Netherlands\\
$^{8}$Max-Planck-Institut Halbleiterlabor, Otto-Hahn-Ring 6, 81739 M\"unchen, Germany\\
$^{9}$Johannes Gutenberg-UniversitŠt, Inst. f. anorganische und analytische Chemie, 55099 Mainz, Germany\\
$^{10}$Department of Physics and Astronomy, University of Leicester, Leicester LE1 7RH, UK}

\begin{document}

\date{Accepted ... . Received ...; in original form ...}


\maketitle

\label{firstpage}

\begin{abstract}

We present the discovery of fast infrared/X-ray correlated variability
in the black-hole transient GX 339-4. The source was observed with
sub-second time resolution simultaneously with VLT/ISAAC and RXTE/PCA
in August 2008, during its persistent low-flux highly variable hard state. The
data show a strong correlated variability, with the infrared emission
lagging the X-ray emission by 100~ms.
The short time delay and the nearly symmetric cross-correlation function, together
with the measured brightness temperature of $\sim 2.5\times 10^6$ K,
indicate that the bright and highly variable infrared emission most likely comes from a jet near the
black hole. Under standard assumptions about jet physics, the measured
time delay can provide us a lower limit of $\Gamma>2$ for the Lorentz
factor of the jet. This suggests that jets from stellar-mass black
holes are at least mildly relativistic near their launching region.
We discuss implications for future applications of this technique.

\end{abstract}


\section{Introduction}

The wealth of multi-wavelength observations of X-ray binaries (XBs) over the
past decade have made clear the ubiquity of jets in these
systems \citep[see][for a review]{fender06}. These jets are thought to
be the origin of the observed radio emission. The soft
X-ray flux is generally believed to come predominantly from
accretion discs around these
compact objects, while the hard X-ray flux is thought to arise from a
hot Comptonizing corona and/or from the jet itself. Recently it
has been shown that also the infrared (IR) emission includes a
substantial contribution from the relativistic jet, in the hard states
of XBs
\citep*[e.g.][]{corbelfender02,malzacetal04,russelletal07}.  Despite
the rapid increase in our phenomenological understanding of jets from
XBs, we still lack a fundamental understanding of how jets
are powered and collimated, or what the bulk and internal properties of the jets are.

High-speed simultaneous optical/X-ray photometry of three accreting black holes (BHs)
opened a new promising window. A complex correlated variability in the optical and X-ray emission
\citep{spruitkanbach02} was seen from XTE J1118+480, while fast
optical photometry of SWIFT J1753.5-0127 \citep{durantetal08} and GX
339--4 \citep[][]{gandhietal08} revealed further
complexity. \citet{malzacetal04} explained the behaviour observed in
XTE J1118+480 through coupling of an optically emitting jet and an
X-ray emitting corona in a common energy reservoir.
An alternative explanation comes from the magnetically driven disc
corona model \citep*{merlonietal00}: magnetic flares happen in an
accretion disc corona where thermal cyclo-synchrotron emission
contributes significantly to the optical emission, while the X-rays
are produced by Comptonization of the soft photons
produced by dissipation in the underlying disc and by the synchrotron
process itself.  The two explanations differ substantially in the predictions at IR wavelengths, where a jet appears as the most probable origin for the emission \citep[e.g.][]{russelletal06}.

The BH candidate GX 339--4 is a recurrent X-ray transient \citep{markert73}. It has been detected as a highly variable source from radio through hard X-rays \citep[see e.g.][and references therein]{makishimaetal86,corbeletal00,coriatetal09}. Optical spectroscopy indicates a mass function of 5.8 $\pm$ 0.5 M$\odot$ and a minimum distance of 6 kpc \citep[][2004]{hynesetal03}. Multiwavelength campaigns clearly reveal a non-thermal contribution to the infrared (IR) emission in the hard state, most probably arising from a compact jet \citep{corbelfender02}. It is the first BH XB for which fast optical/X-ray correlated variability was observed \citep*{motchetal82}. 

Past variability studies on timescales of several seconds have
been used to suggest a jet origin for IR emission from XBs \citep[e.g.][]{hynesetal03,eikenberryetal08}.  In this
Letter, we report on the first simultaneous fast (i.e. sub-second)
timing IR/X-ray observations of GX 339--4, during its 2008 low
luminosity hard state.
\vspace{-3mm}
\section{Observations}\label{observations}

\subsection{Infrared data}

We observed GX~339-4 from ESO's Paranal Observatory on 18 August 2008.
We obtained fast $K_s$-band photometry with the Infrared Spectrometer And Array Camera (ISAAC) \citep{moorwoodetal98}
mounted on the 8.2-m UT1/Antu telescope. The 23$\arcsec\times$23$\arcsec$ window
used encompassed the target, a bright `reference' star
(K$_{\rm S}$=9.5)
located 13.6 arcsec south of our
target and a fainter `comparison' star (K$_{\rm S}$=12.8)
8.9 arcsec north-east of GX339-4.

\begin{figure}
\begin{center}
\includegraphics[width=83mm]{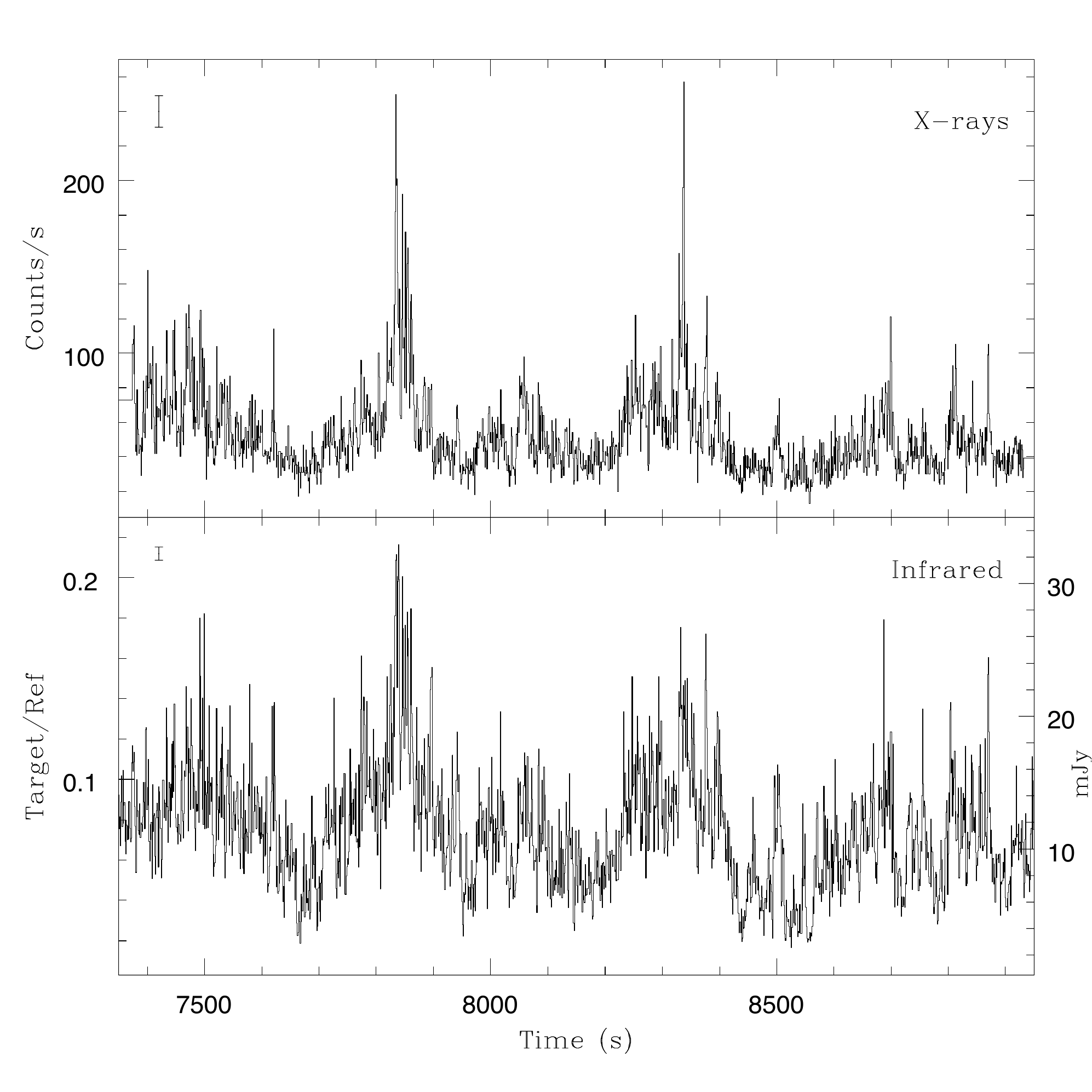}
\end{center}
 \caption{{\it Top panel:} A sample of the X-ray light curve of GX 339--4, obtained with the PCA onboard RXTE. The data are background subtracted, in the 2-15 keV energy range, at 1-second time resolution. {\it Bottom panel:} The simultaneous IR light curve, obtained with ISAAC. We show the ratio between the source (average $4.4\times10^5$ counts/s) and the reference-star ($6\times10^6$ counts/s) count rates in the K$_S$ filter, at 1-second time resolution. The right ordinates show the de-reddened flux. \new{We show the typical error bars in the top-left corner of each panel.}}
\label{lcurves}
\end{figure}

We used the ``FastJitter mode'' with a time resolution of 62.5
ms. This generated cubes of data with 2500 images apiece, and with a
3~s deadtime between cubes. The ULTRACAM pipeline\footnote{We thank
Tom Marsh for the use of the ULTRACAM pipeline software
(http://deneb.astro.warwick.ac.uk/phsaap/software/).} was used for the
data reduction, after applying a barycenter correction for Earth
motion. We performed fixed-aperture photometry of the three sources (target,
reference and comparison stars) and used the bright reference star for
relative photometry of the target and comparison stars. The positions
of the aperture regions around the target and the comparison star were
linked to the position of the bright reference star to allow for image
motion and were updated at each time step. The atmospheric conditions
were good and the resulting light curve for the comparison star was
consistent with a constant, as expected. By combining all 250000
images, we estimate a de-reddened \citep[$A_V$=3.9, $A_K=0.114\times
A_V$=0.445--][]{cardellietal89} average magnitude of K$_{\rm S}$=12.0$\pm$0.2
for GX~339-4, which corresponds to an average flux of $F\sim1.5 \times
10^{-11}$~erg~s$^{-1}$~cm$^{-2}$.
A sample of the highly variable light curve for GX~339-4 is shown in the bottom panel of Fig.~\ref{lcurves}.

\subsection{X-ray data}

Simultaneously with the IR observations, GX 339--4 was observed with
the Proportional Counter Array (PCA) onboard the {\it Rossi X-ray
Timing Explorer (RXTE)}. Two proportional counter units (PCUs) were
active during the whole observation. The X-ray data span three
consecutive satellite orbits, for a total exposure of 4.6 ksec.
The {\tt Binned Mode} (8~ms time resolution)
was used for this analysis, using the 2-15 keV energy range
(channels 0-35). The barycenter correction for Earth and satellite motion was
applied.  Standard HEADAS 6.5.1 tools were used for data reduction. In
the upper panel of Fig.~\ref{lcurves} we show a sample of the light
curve, corresponding to the second {\it RXTE} orbit. Spectral fitting
with a power-law with photon index 1.6 results in a 2--10 keV unabsorbed flux of
$F_{\rm X}\sim1.4\times 10^{-10}$ erg~s$^{-1}$~cm$^{-2}$. 

\section{Cross-correlation function (CCF)}

Both datasets have an absolute time accuracy better than the time
resolution used here: ISAAC data have a timing accuracy of about 10 ms
(the readout time), while RXTE data have a timing accuracy of 2.5
$\mu$s \citep{jahodaetal06}.

\begin{figure}
\begin{center}
\includegraphics[width=83mm]{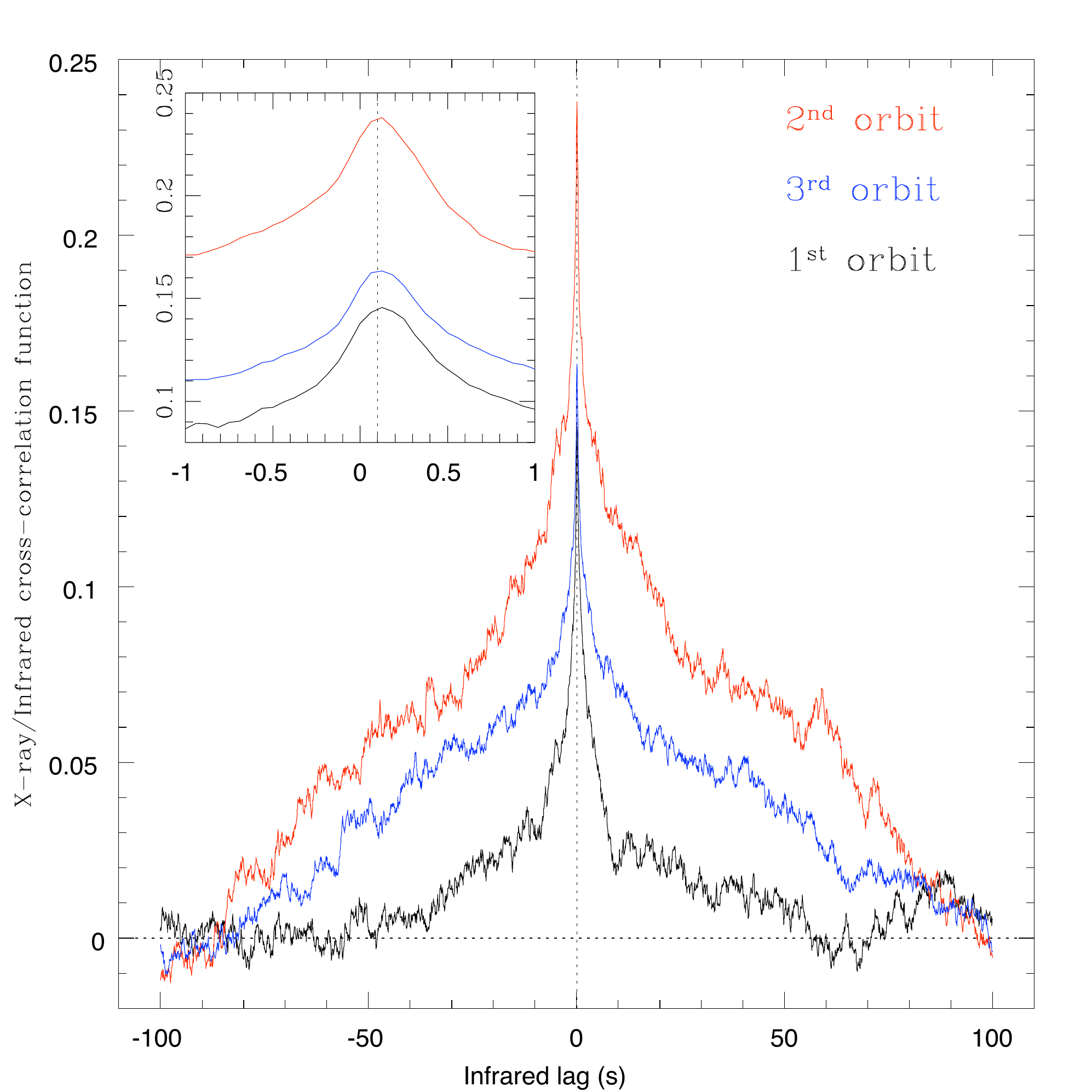}
\end{center}
 \caption{Cross-correlations of the X-ray and IR light curves of GX 339--4 (positive lags mean IR lags the X-rays). A strong, nearly symmetric correlation is evident in all the three time intervals, corresponding to different RXTE orbits. In the inset we show a zoom of the peaks, showing the IR delay of $\sim$100~ms with respect to the X-rays. The inset also shows a slight asymmetry toward positive delays.}
\label{ccf}
\end{figure}

From Fig.~\ref{lcurves} a strong correlation between X-ray and IR flux is evident.
Both long, smooth variability and short, sharper
flares appear with similar relative amplitude in the two energy
bands. In order to measure any time delay, we calculated a
CCF for each of the three {\it RXTE} orbits, without applying any
de-trending procedure. The results are shown in Fig.~\ref{ccf}. The
strong correlation is confirmed. The CCF appears highly
symmetric and relatively stable over the three time intervals, with
the change in amplitude simply reflecting the different variability
amplitude in the light curves themselves. In the inset, we show a zoom
on the peak of the CCF, which shows how the IR emission lags 
the X-rays by 0.1 seconds, \new{to which we associate an uncertainty of 30$\%$ (which includes systematics).}

\section{Discussion}

The main result of our work is the discovery of a strong correlation
between the IR and the X-ray variability in GX 339--4. The fact that
the CCF is nearly symmetric and peaks at 100~ms rules out a
reprocessing origin for the IR variability. If the IR radiation arose
from reprocessing of X-rays by the outer disk, the short time delay
would imply a highly inclined disk. This would produce an highly asymmetric
CCF, with a tail at long lags \citep{obrienetal02}.

\begin{figure}
\begin{center}
\includegraphics[width=83mm]{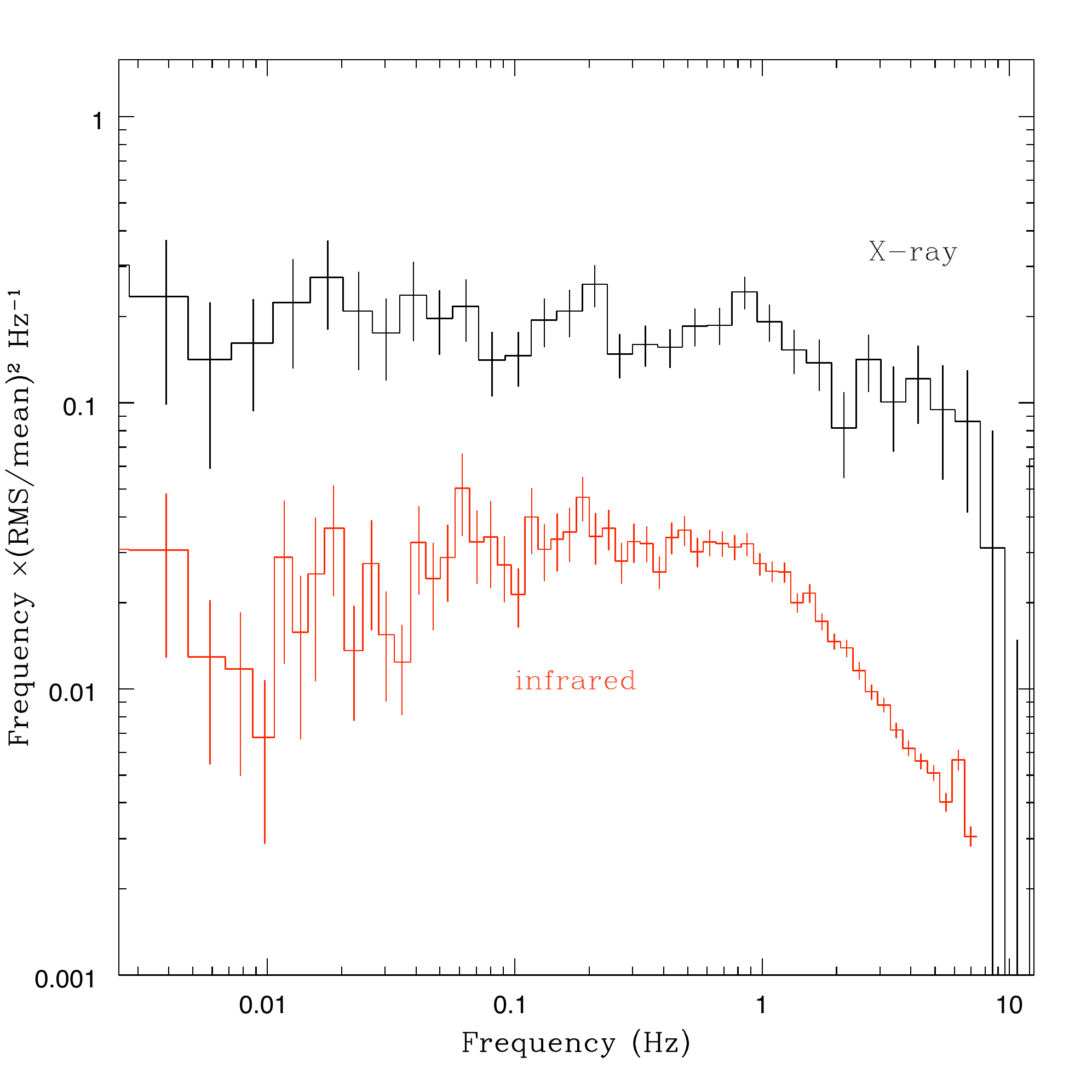}
\end{center}
 \caption{X-ray (2-15 keV) power spectrum of the second RXTE orbit (upper curve), together with the power spectrum of the simultaneous IR light curve (lower curve). The Poissonian noise has been subtracted from both spectra. The peak at $\sim$6 Hz in the IR spectrum is instrumental. The high-frequency portion of the IR spectrum has yet un-modeled systematics, which however do not affect the results presented here.}
\label{pds}
\end{figure}

Additionally, power spectral analysis shows significant IR variability
(at least 5\% fractional rms, see Fig.~\ref{pds}) on timescales of $\sim$200~ms or
shorter, which sets an upper limit of $\sim 6 \times$ 10$^9$ cm to the
radius of the IR-emitting region. From (5\% of) the observed IR average
flux of $F\sim 1.5\times10^{-11}$ erg s$^{-1}$ cm$^{-2}$, we derive
a minimum brightness temperature of $\sim2.5\,\times 10^6$ K. Optically
thick thermal emission of the derived size and temperature would result in a 2--10 keV flux in
excess of $10^{-5}$ erg s$^{-1}$ cm$^{-2}$, which is not observed in the
data. These values represent very conservative estimates: a smaller
region emitting the IR radiation would result in a higher brightness
temperature, which would in turn result in a higher expected X-ray
luminosity.
With similar arguments we exclude thermal Bremsstrahlung emission. 
The existence of an IR lag is also inconsistent with the 
magnetic corona model \citep{merlonietal00}, in which the same population of electrons produces the IR synchrotron emission and the X-ray Compton emission.
We conclude that the most plausible origin for the observed IR variability is synchrotron emission from the inner jet.\footnote{This \new{is confirmed} by nearly-simultaneous optical and IR observations, obtained while the source was in the same low-luminosity state. Those data (Lewis et al., in prep.) show a flat or inverted spectrum (\new{inconsistent} with thermal emission from a disc or companion star), and a long-timescale ($\sim$minutes) variability stronger in IR than in optical.}

This result is a new, independent strong indication that jet synchrotron emission
contributes significantly to the IR radiation in this source. This is
the first time that hard-state, compact jet emission has been securely
identified to vary on sub-second timescales in an XB, although
variability on similar dynamical timescales \new{t$_{Dyn}$  (i.e., scaled to mass)} had been already observed
in Active Galactic Nuclei \citep[e.g.][]{schodeletal07}. These data
thus represent a further step forward towards a full unification of
the accretion/ejection process over a broad range of black-hole
masses.

\subsection{Emitting regions and jet speed}

Our data strongly suggest that the variable IR emission comes from
the jet, although we cannot conclude whether it is optically--thick or
--thin synchrotron. The X-ray emission is usually interpreted as
Comptonized radiation from energetic plasma in the very inner regions
of the accretion flow, although the actual emitting region is still an
open issue \citep*[either a corona or the base of the jet itself, for
a discussion see][Malzac et al. 2009]{markoffetal05,maccarone05}. Here
we discuss the four possible scenarios.

{\it 1) IR: optically thick - X-ray: inflow emission:}
the observed time delay between the IR and the X-ray variability gives
an upper limit (given the unknown time for the ejection to take place)
to the travel time of the variability -- thus presumably the
matter -- along the jet. Given a measure of the jet elongation we could
estimate the jet speed. Such a measure is not available for GX~339-4;
however, a jet elongation measurement has been reported from 8.4 GHz
observations of another BH XB, Cyg~X-1
\citep{stirlingetal01}. Within the standard model for compact jets
\citep[][-- BK79]{BK79}, the distance from the black hole of an
emitting region in the jet is a function of the observed wavelength,
the viewing angle, and the jet physical properties.

Assuming that the main physical properties of the jet do not change,
we can rescale the jet elongation measured at radio wavelengths in the
BH Cyg X-1 down to the IR wavelengths, obtaining a measurement
of the distance of the IR-emitting region in the jet from the black
hole in GX 339-4. \citet{HM04} showed that the small observed scatter
in the radio/X-ray fluxes relation \citep*{galloetal03} implies that
all stellar-mass BHs have very similar jet velocities, with a
3-sigma spread of $\Delta(\beta\Gamma)<1.6$. \new{(We note however that there is an increasing population of radio-quiet sources, whose nature is not yet understood \citep[see e.g.][Soleri et al. submitted.; and references therein]{gallo07,casellapeer09}. Their inclusion in the radio/X-ray flux correlation increases the scatter of the correlation itself, so the conclusions drawn by \citet{HM04} do not necessarily hold any longer.)}

We can use Eq.~28 of BK79 to obtain the GX 339--4 jet elongation in IR, 
and thus the jet speed. We calculate the jet speed for a
total of $10^5$ sets of parameters, with each parameter randomly
chosen within its measured permitted range. For Cyg X--1, we use a
distance of 2.0$\pm$0.2 kpc \citep{distancecgx1} and a source
inclination over the 20$^\circ$-70$^\circ$ range
\citep{dolan92,ziolkowski05}. We use a projected jet elongation of
L$_{Jcyg} = 2.6\times10^{14}\times$D$_{kpc}$ cm (where $D_{\rm kpc}$ is
the distance of Cyg~X-1 in kpc), to which we attribute a 50\%
uncertainty. For GX 339-4, we used
the minimum distance of 6 kpc \citep[][2004]{hynesetal03} and a range of
inclination angles 15$^\circ$-60$^\circ$
\citep{cowleyetal02,milleretal08}.

We use the X-ray luminosity as a tracer of jet kinetic power, under
the additional assumption that the jet power follows the $P_J \propto
L_{\rm X}^{0.5}$ relation \citep*{fenderetal03}.  We estimate the X-ray
(2--10 keV) fluxes of the two sources through spectral fitting of the
RXTE data simultaneous to the IR (GX 339--4, $F_{\rm X}=1.4\times
10^{-10}$ erg cm$^{-2}$ s$^{-1}$) and radio (Cyg X--1, $F_{\rm
X}=9\times 10^{-9}$ erg cm$^{-2}$ s$^{-1}$) observations, and we
attribute a 50\% uncertainty to both values, to account for the
uncertainties in the $P_J - L_X$ scaling.

We obtain a \new{3.3}-$\sigma$ lower limit of $\Gamma > 2$.
It is important to remember that the reliability
of this estimate depends on the key assumptions made in BK79 being
true -- the most important of these, that particle acceleration along
the jet is continuous and counterbalances energy losses in order to
preserve the flat spectrum -- is largely untested \citep[see e.g.][]{kaiser06,peercasella09}. 
However, we have chosen
very conservative ranges for all parameters. The lowest Lorentz
factors (2--3) are obtained only for small angles of Cyg X--1, for the minimum Cyg X--1
jet elongation within the allowed range and for the minimum GX 339--4
distance.  We conclude that,
if the IR synchrotron emission is optically-thick, these data suggest that the jets from accreting stellar-mass
BHs are at least mildly relativistic, also in their common
low/hard state. If, as is widely suggested, the jet speed corresponds
to the escape speed at the launch point, this might imply that the jet
is launched from a region very close to the black hole itself.

The obtained values of $\Gamma$ might allow us further
considerations. Assuming a random distribution of jet inclination
angles $\theta$ for the known population of stellar-mass BHs
(i.e., P$_{\theta} \propto$ sin$\theta$), all with a Lorentz factor of
2, about 57\% of them would appear de-boosted. This percentage
increases up to 70\% for $\Gamma=3$, and up to $\sim$77\% for
$\Gamma=4$. Thus our jet speed estimate suggests that more than half
of the jet luminosities measured from BHs in our
Galaxy might have been underestimated. If this is true, there might be
a population of BH binaries with very high radio to X-ray
ratios, which will be revealed in our and other galaxies by
 future radio all-sky monitors and large radio telescopes.
The obtained estimate for $\Gamma$ has several
caveats, or at least large uncertainties. Future monitoring observations
with the same technique will allow to refine this measure, studying the relative dependency of
this quantity with the varying accretion rate or total luminosity

{\it 2) IR: optically thin - X-ray: inflow emission:} this scenario is actually twofold: the IR optically thin emission
could originate at the first shock in the jet, or further away after cooling. In the first case, the observed time delay
would measure the ejection and first acceleration timescale. In the second case, since we expect the spectral break to be around the IR wavelengths, we can approximate the IR radiation as if it was
optically thick, and use Eq.~28 of BK79 as above. Thus, the lines of reasoning described in the previous scenarios hold, as well as the lower limit to the jet speed.

{\it 3) IR: optically thick - X-ray: jet emission:} 
the reasoning described in the first scenario, as well as the resulting lower limit for $\Gamma$, still hold, provided that the X-ray emission is not boosted \citep[for a discussion about the evidence for this be the case, see][]{HM04}.

{\it 4) IR: optically thin - X-ray: jet emission:} the electron
populations emitting at IR and X-ray wavelengths cannot be co-located,
since we observe a delay, which here must represent a cooling
time. If we assume that the X-ray emission gives us a measure of the characteristic energy of the
emitting electrons, we can place an upper limit on the
magnetic field intensity in the jet \citep[see e.g.][for a use of this
method in blazars]{takahashi96}. We obtain a unique solution for the
system, with a magnetic field intensity of B$\sim 10^4$ Gauss, an
initial Lorentz factor of the electrons of $\gamma \sim 7\times 10^3$,
which becomes $\gamma \sim 50$ after they cool down as to emit in IR.
Electron re-acceleration \citep[e.g.][]{jamiletal09} would act as to
counter balance the radiative losses, resulting in higher values for
the magnetic field intensity.

\subsection{Comparison with optical/X-ray CCFs}

The main difference between this IR/X-ray CCF and the published
optical/X-ray CCFs (for this as well for other BHs) is the longer IR
wavelength itself, which allows us to put strong constraints on the
emission processes. In the hard state of BHs, the jet is expected to dominate the IR emission \citep[][see also $^2$]{russelletal06}. In fact, as we discussed in the previous section, these
data allow us to rule out any thermal origin for the observed variability. 
To put similar constraints using optical data, a much higher luminosity (or faster variability) is needed \citep[see e.g.][]{motchetal82}.

Our CCF is very different from the one observed between the
optical and the X-ray variability \citep{gandhietal08}. Namely, the
sharp positive peak in our dataset is not preceded nor followed by any
dip. This might be due to the additional presence of a large
long-timescale variability, which buries the short-timescale structure
of the IR CCF. In fact we did not apply any detrending procedure to
our data, while this has been partially done for obtaining the optical
CCF (Gandhi et al., in prep.). That this can be the reason for the
observed differences seems to be suggested also by the fact that
the IR CCF with the lowest amplitude among the three orbits
(black curve in Fig.~\ref{ccf}) appears steeper towards the positive time
delays, similarly to the optical/X-ray CCF of this source. On the
other hand, it might be that the the dips observed in the optical CCFs arise from a
spectral component which does not contribute much at IR wavelengths. A
detailed study of this will be presented in future work.

The two CCFs differ also in the time delay measured at their peak: the
optical/X-ray CCF peaks at $\sim 150$~ms, a 50-ms longer delay than
IR \new{(although we note that, depending on the uncertainty on the measured optical delay, the two delays might be marginally consistent within the errors)}. Given that the optical emission should come from closer to the
black hole than the IR emission, this result is contrary to simple
expectations.  The two observations occured at similar X-ray
luminosities, casting doubt on jet power variation as a reason for the
longer optical than IR delay. However, the optical data were
acquired after the decay of a bright outburst, while our IR data were
acquired at the end of the rise of a weak outburst, suggesting a
hysteresis effect might play a role
\citep*{vadawaleetal03,fenderbellonigallo04,russelletal07b}. In
particular, during/after the decay of an outburst the emission at a
given wavelength might happen at larger distances from the black hole,
because of the smaller amount of previously ejected matter that the
jet encounters. Future {\it simultaneous} IR, optical and X-ray observations, at
high-time resolution, will allow us to unveil some of these important
unknowns, thus allowing us to test some of the physical assumptions
which are now behind this method.

\section*{Acknowledgments}

We thank the RXTE and VLT schedule planners for their successful
efforts in scheduling these simultaneous observations. PC thanks ESO
for the financial support, and the staff in Santiago for the friendly
hospitality. PC thanks P. Gandhi, F. Panessa and G. Ponti for useful discussions. This work was partially supported by an NWO Spinoza grant
to M. van der Klis. PC acknowledges funding via a EU Marie Curie
Intra-European Fellowship under contract no. 2009-237722.
DMR acknowledges support from a NWO Veni Fellowship.
TB thanks ASI/INAF for support through grant I/088/06/0. 
TJM and TB thank the EU FP7 for support through grant number ITN 215212 "Black Hole Universe".


\label{lastpage}


\begin{thebibliography}{99}

\bibitem[\protect\citeauthoryear{Blandford \& K\"onigl}{1979}]{BK79} Blandford R. D. \& K\"onigl A., 1979, ApJ, 232, 34

\bibitem[\protect\citeauthoryear{Cardelli, Clayton \& Mathis}{1989}]{cardellietal89} Cardelli J., Clayton G., Mathis J., 1989, ApJ, 345, 245

\bibitem[\protect\citeauthoryear{Casella \& Pe'er}{2009}]{casellapeer09} Casella P., Pe'er A., 2009, ApJL, 703, L63

\bibitem[\protect\citeauthoryear{Corbel et al.}{2000}]{corbeletal00} Corbel S., Fender R. P., Tzioumis A. K., Nowak M., McIntyre V., Durouchoux P., Sood R., 2000, A\&A, 359, 251

\bibitem[\protect\citeauthoryear{Corbel \& Fender}{2002}]{corbelfender02} Corbel S., Fender R. P., 2002, ApJ, 573, 35

\bibitem[\protect\citeauthoryear{Coriat et al.}{2009}]{coriatetal09} Coriat M., Corbel S., Buxton M. M., Bailyn C. D., Tomsick J. A., K\"ording E., Kalemci E., 2009, MNRAS, tmp1311

\bibitem[\protect\citeauthoryear{Cowley et al.}{2002}]{cowleyetal02} Cowley A. P.,  Schmidtke P. C., Hutchings J. B., Crampton D.,2002, AJ, 123, 1741

\bibitem[\protect\citeauthoryear{Dolan}{1992}]{dolan92} Dolan J. F., 1992, ApJ, 384, 249

\bibitem[\protect\citeauthoryear{Durant et al.}{2008}]{durantetal08} Durant M., Gandhi P., Shahbaz T., Fabian A. P., Miller J., Dhillon V. S., Marsh T. R., 2008, ApJ, 682, 45

\bibitem[\protect\citeauthoryear{Eikenberry et al.}{2008}]{eikenberryetal08} Eikenberry S. S., Patel S. G., Rothstein D. M., Remillard R., Pooley G. G., Morgan E. H., 2008, ApJ, 678, 369

\bibitem[\protect\citeauthoryear{Fender, Gallo \& Jonker}{Fender et al.}{2003}]{fenderetal03} Fender R.P., Gallo E., Jonker P.G., 2003, MNRAS, 343, 99

\bibitem[\protect\citeauthoryear{Fender, Belloni \& Gallo}{Fender et al.}{2004}]{fenderbellonigallo04} Fender R. P., Belloni T., Gallo E., 2004, MNRAS, 355, 1105

\bibitem[\protect\citeauthoryear{Fender}{2006}]{fender06} Fender R. P., 2006, Jets from X-ray binaries, in Compact Stellar X-Ray Sources, ed. W. H. G. Lewin \& M. van der Klis (Cambridge: Cambridge University Press), 381-419


\bibitem[\protect\citeauthoryear{Gallo, Fender \& Pooley}{Gallo et al.}{2003}]{galloetal03} Gallo E., Fender R., Pooley G., 2003, MNRAS, 344, 60

\bibitem[\protect\citeauthoryear{Gallo}{2007}]{gallo07} Gallo E., 2007, AIP Conference Proceedings, 914, 715

\bibitem[\protect\citeauthoryear{Gandhi et al.}{2008}]{gandhietal08} Gandhi P., Makishima K., Durant M., Fabian A. C., Dhillon V. S., Marsh T. R., Miller J. M., Shahbaz T., Spruit H. C., 2008, MNRAS, 390, L29

\bibitem[\protect\citeauthoryear{Gierlinski et al.}{1999}]{distancecgx1} Gierlinski M., Zdziarski A. A., Poutanen J., Coppi P. S., Ebisawa K., Johnson W. N., 1999, MNRAS, 309, 496

\bibitem[\protect\citeauthoryear{Heinz \& Merloni}{2004}]{HM04} Heinz S., Merloni A., 2004, MNRAS, 355, L1



\bibitem[\protect\citeauthoryear{Hynes et al.}{2003}]{hynesetal03} Hynes R. I., Steeghs D., Casares J., Charles P. A., O'Brien K., 2003, ApJL, 583, L95

\bibitem[\protect\citeauthoryear{Hynes et al.}{2004}]{hynesetal04} Hynes R. I., Steeghs D., Casares J., Charles P. A., O'Brien K., 2004, ApJ, 609, 317

\bibitem[\protect\citeauthoryear{Jahoda et al.}{2006}]{jahodaetal06} Jahoda K., Markwardt C.~B., Radeva Y., Rots A.~H., Stark M.~J., Swank J.~H., Strohmayer T.~E., Zhang W., 2006, ApJS, 163, 401

\bibitem[\protect\citeauthoryear{Jamil et al.}{2009}]{jamiletal09} Jamil O., Fender R., Kaiser C., 2009, MNRAS, tmp1543

\bibitem[\protect\citeauthoryear{Kaiser}{2006}]{kaiser06} Kaiser C. R., 2006, MNRAS, 367, 1083

\bibitem[\protect\citeauthoryear{Maccarone}{2005}]{maccarone05} Maccarone T. J., 2005, MNRAS, 360, 68

\bibitem[\protect\citeauthoryear{Malzac, Merloni \& Fabian}{Malzac et al.}{2004}]{malzacetal04} Malzac J., Merloni A., Fabian A., 2004, MNRAS, 351, 253

\bibitem[\protect\citeauthoryear{Makishima et al.}{1986}]{makishimaetal86} Makishima K., Maejima Y., Mitsuda K., Bradt H. V., Remillard R. A., Tuohy I. R., Hoshi R., Nakagawa M., 1986, ApJ, 308, 635

\bibitem[\protect\citeauthoryear{Markert et al.}{1973}]{markert73} Markert T.~H., Canizares C. R., Clark G. W., Lewin W. H. G., Schnopper H. W., Sprott G. F., 1973, ApJ, 184, L67

\bibitem[\protect\citeauthoryear{Markoff, Nowak \& Wilms}{Markoff et al.}{2005}]{markoffetal05} Markoff S., Nowak M. A., Wilms J., 2005, ApJ, 635, 1203

\bibitem[\protect\citeauthoryear{Merloni, Di Matteo \& Fabian}{Merloni et al.}{2000}]{merlonietal00} Merloni A., Di Matteo T., Fabian A. C., 2000, MNRAS, 318, 15

\bibitem[\protect\citeauthoryear{Miller et al.}{2008}]{milleretal08} Miller J. M., Reynolds C. S., Fabian A. C., Cackett E. M., Miniutti G., Raymond J., Steeghs D., Reis R., Homan J., 2008, ApJ, 679, 113

\bibitem[\protect\citeauthoryear{Moorwood et al.}{1998}]{moorwoodetal98} Moorwood A., et al., 1998, Messenger, 94, 7

\bibitem[\protect\citeauthoryear{Motch, Ilovaisky \& Chevalier}{Motch et al.}{1982}]{motchetal82} Motch C., Ilovaisky S. A., Chevalier C., 1982, A\&A, 109, 1

\bibitem[\protect\citeauthoryear{O'Brien et al.}{2002}]{obrienetal02} O' Brien K., Horne K., Hynes R. I., Chen W., Haswell C. A., Still M. D., 2002, MNRAS, 334, 426

\bibitem[\protect\citeauthoryear{Pe'er \& Casella}{2009}]{peercasella09} Pe'er A., Casella P., 2009, ApJ, 699, 1919

\bibitem[\protect\citeauthoryear{Russell et al.}{2006}]{russelletal06} Russell D., Fender R., Hynes R., Brocksopp C., Homan J., Jonker P., Buxton M., 2006, MNRAS, 371, 1334

\bibitem[\protect\citeauthoryear{Russell, Fender \& Jonker}{Russell et al.}{2007}]{russelletal07} Russell D., Fender R., Jonker P., 2007, MNRAS, 379, 1108

\bibitem[\protect\citeauthoryear{Russell et al.}{2007}]{russelletal07b} Russell D., Maccarone T., K\"ording E., Homan J., 2007, MNRAS, 379, 1401

\bibitem[\protect\citeauthoryear{Sch\"odel et al.}{2007}]{schodeletal07} Sch\"odel R., Krips M., Markoff S., Neri R., Eckart A., 2007, A\&A, 463, 551

\bibitem[\protect\citeauthoryear{Spruit \& Kanbach}{2002}]{spruitkanbach02} Spruit H. C., Kanbach G., 2002, A\&A, 391, 225

\bibitem[\protect\citeauthoryear{Stirling et al.}{2001}]{stirlingetal01} Stirling A. M., Spencer R. E., de la Force C. J., Garrett M. A., Fender R. P., Ogley R. N., 2001, MNRAS, 327, 1273

\bibitem[\protect\citeauthoryear{Takahashi et al.}{1996}]{takahashi96} Takahashi T., Tashiro M., Madejski G., Kubo H., Kamae T., Kataoka J., Kii T., Makino F., Makishima K., Yamasaki N., 1996, ApJL, 470, L89

\bibitem[\protect\citeauthoryear{Vadawale et al.}{2003}]{vadawaleetal03} Vadawale S., Rao A., Naik S., Yadav J., Ishwara-Chandra C., Pramesh Rao A., Pooley G., 2003, ApJ, 597, 1023

\bibitem[\protect\citeauthoryear{Ziolkowski}{2005}]{ziolkowski05} Ziolkowski J., 2005, MNRAS, 358, 851



\end{thebibliography}
\end{document}